# Design Artifact's, Design Principles, Problems, Goals and Importance

Zeeshan Ahmed

Vienna University of Technology,
Address: Getreidemarkt 9/307 1060,
Vienna, Austria
Mobile: 004369981302854,
Email: zeeshan.ahmed@tuwien.ac.at
URL: www.tuwien.ac.at &
www.zeeshanahmed.bravehost.com

Sudhir Kumar Ganti

Securitas Direct AB
Address: härds väg 6B, lgh 847 21367
Malmo, Sweden
Mobile: 0046735549798,
Email: sudhir.ganti@securitas-direct.com
URL: www.securitas-direct.com

Hans Kyhlbäck

Department of interaction and system design, Blekinge Institute of Technology, School of Engineering
Address: PO Box 520, SE-372 25 Ronneby, Sweden
Phone: +46 457 38 58 63 & Email: hky@bth.se & URL: www.bth.se/tek

*Abstract*— **Designing human computer interaction interface is an important and a complex task, but it could be simplified by decomposing task into subcomponents and maintaining relationships among those subcomponents. Task decomposition is a structured approach, applicable in both Software Engineering and Human Computer Interaction (HCI) fields depending on specific processes and design artifacts. Using design artifacts applications could be made for analysis and design by making the hand draw sketches to provide high level of logical design based on user requirements, usage scenarios and essential use cases. To design hand draw sketches there are some strategies to be followed .i.e., planning, sequential work flow, and levels of details. In this research paper we are presenting design artifacts, goals, principles, guidelines and currently faced problems to human computer interaction design community. Moreover in the end concluded with assessed observations in a case study**

## I. INTRODUCTION

THIS report focuses on the design, design Goals, some major principles, some currently faced problems and importance of the Design Artifacts. "Design" as a verb refers to the process of originating and developing a plan for a new object .i.e., machine, building, product etc, As a noun, "design" is used both for the final plan or proposal .i.e., a drawing, model or the produced object. The design of graphical user interface of especially big software is not only expensive and time-consuming, but it is also critical for effective system performance. Design is driven by the requirements like what artifacts is for and how is to be implemented. Design represents the artifacts as story boards, screen sketches, tasks flow diagrams and finally in executable prototypes[8]. Moreover, Design keeps the understanding of the role of the task for which the interface will be used and the work environment in which the interface will be applied. Experience various approaches to the identification of critical factors influencing interface design and development.

Artifact document is the design decision and the architecture leading to the required design. Artifact is mainly divided into two categories .i.e., requirement specification and system architecture [1]. Requirements specification is the description of functional and non functional requirements whereas system architecture includes context, archetypes, structure and design decisions. Context is the interfaces defined to the external entities, archetypes are the core abstractions on which the system is build, structure is the component that constitutes the system and finally the design decisions are the rules, constraints and transformation of the system architecture.

In this research article we first present an introduction to Human Computer Interaction (HCI), briefly describing its three main design ideologies in section II. Then we provide the information about design principles in section III, design patterns in section IV, design guild lines in section V, basic design goals in section VI, currently faced major design based problems in section VII and in the end we have presented a real time case study describing the whole software application development process emphasizing on the design of graphical user interface in section VIII.

## II. HUMAN COMPUTER INTERACTION (HCI)

Human Computer Interaction (HCI) is the study of design, evaluation and implementation of interactive computing systems for human use [7]. HCI mainly consists of three ideologies .i.e., Design idea, design activity and design learning. Design Idea is the tradition of arts where as design activity is the action performed by artifact(s). Design learning is the main and important ideology comprises of three important elements .i.e., iDeas notebook, the iDeas blog, and iDeas wall. [10].

### A. iDeas notebook

iDeas notebook extends the traditional notebook design by merging the physical and electronic input in to sketches. The main activities of iDeas notebook are electronic file based activities which are automatically created by notebook page photographs and while retaining the physical aspects of the idea log. There are some advantages of iDeas notebook that they are rapidly capturing rich amounts of data, ready at hand and familiar interaction, permitting designers to focus on tasks rather than tools, and accompanying users in the field and wherever, whenever design happens.

*B. iDeas blog*

iDeas blog serves as a digital store of collected and generated information. There are two sources of inspiration for iDeas blog .i.e., traditional blog and shared electronic portfolios. Traditional blog, are primarily text based and require lots of interactions to add visual information where as shared electronic portfolios follows highly formalized way to explicitly support visual and textual information. The iDeas blog is the extension of lightweight, automatic integration and archival of iDeas notebook inputs.

*C. iDeas wall*

iDeas wall provides an interactive surface to present and create ideas and general purpose contents. It is a vertical display surface with direct manipulation capability which can afford collocated group interactions, including the presentation interaction style and the whiteboard interaction style of brainstorming sessions. The iDeas wall provides three methods for users to create and import content: they can sketch and write on the wall as they would on a whiteboard; they can import content from the iDeas blog; or they can bring up an iDeas notebook page directly by using the pen as a command device. Content created on the wall is saved to the iDeas blog.

III. DESIGN PRINCIPLES

There are four main design principles [9] [11].

*A. Cooperation*

Cooperation plays a vital role in software project development. The important and primitive principle of design process is the cooperation between both developers and the end users. Because in the design process with respect to the participatory design point of view there exists an uncommon principle .i.e., presenting the same issues with completely different perspectives and dimensions.

*B. Experimentation*

Generally experimentation is performed in the middle of recently acquired possibilities and the currently existing conditions. To assure that the present conditions are in conjunction with new ideas and supported by two primitive principles .i.e., concretization and contextualization of design, Principles are in associated with the above mentioned visions performing experiments with visions and hand on experience.

*C. Contextualization*

Design hooks its initial point with a particular configuration in which new computer based applications put into practice. Participatory design emphasizes on situations based on the implementation of iterative designs. The design composition of use is tied up with numerous social and technical issues. Generally participatory design of the development will specifically includes different kinds of participants i.e. Users, Managers and the design developers.

*D. Iteration*

In design process, we should hang on to some issues which are not yet revealed, which are visioning the future product from design point of view and the construction of work from use point of view. But participatory design puts a controversial statement in accomplishing the same by making use of artifacts i.e. Prototype. Designers with cooperation will make use of the artifacts as a source for delegation of work. Participatory design also ends up with a controversial statement for trivial division of work in the process of development, which pleads overlap among the members of analysis, design and realization groups.

IV. DESIGN PATTERNS

Like software engineering design patterns there are also some graphical user interface design patterns [12] .i.e., Window Per Task, Interaction Style, Explorable Interface, Conversational Text, Selection, Form, Direct Manipulation, Limited Selection Size, Ephemeral Feedback, Disabled Irrelevant Things, Supplementary Window, Step-by-Step Instructions [13].

*A. Window Per Task*

This design pattern specifies organization of the user interface into windows dedicated into different tasks. This organizes all the user interface components.[2]

*B. Interaction Style*

This pattern helps in selecting the primary way in which users interact with windows. This pattern is also identifies the interaction style for the users in the form of Selection, Direct Manipulation. [4]

*C. Explorable Interface*

This pattern provides the guidance on minimizing the cost of user's mistakes by providing the information about exceptions and errors. [4]

*D. Conversational Text*

This pattern provides information about designing a GUI to accept commands in the form of textual input. [13]

*E. Selection*

This pattern describes a style of user interaction where the user chooses selection from the list. [6]

*F. Form*

This pattern allows a user to input discrete structured data as in put into the GUI. [13]

*G. Direct Manipulation*

This pattern provides guidance on how to structure user

interactions. [6]

*H. Limited Selection Size*

This pattern provides the guidance on how to structure sets of selection. [6]

*I. Ephemeral Feedback*

This pattern without interfering with the natural flow of applications provides feedback to users about the current status of the work. [13]

*J. Disabled Irrelevant Things*

This pattern provides guidance on how to hide or disable GUI elements that are not relevant to current context. [13]

*K. Supplementary Windows*

This pattern provides information about supplementary window. [6]

*L. Step-by-Step Instructions*

This pattern helps the user in performing the required task by performing certain actions in a sequence using applications. [13]

## V. DESIGN GUIDELINES

A successful design interface can be implementable using the following guidelines .i.e.,
- Design should be presentable
- Criteria / principles should be applied
- Prepared according to the project proposal
- Specified according to the requirements
- Should be evaluated
- Assessed by mental work load
- Design mock-ups should be implemented
- A safe analysis should be performed
- Should be flexible enough to adopt rapidly prototyped changes and modifications
- Design should be iterative.
- Use case modeling[1] should be used with the identification of user interface elements

## VI. BASIC DESIGN GOALS

A successful design interface has three main primary design goals .i.e., high resolution, clean screen and fluid interaction.

*A. High Resolution*

An interactive design must display required quantitative material including images and application windows at an appropriate resolution for the user standing at the board (a resolution comparable to workstation monitors). To provide a smooth wall like environment, it needs to be flat, continuous surfaced and without physical seam interruption. In the future most probably large high resolution based flat panels will be available.

---
[1] UML can be used for use case modeling [3]

*B. Clean Screen*

There is a striking difference between the visual look of a paper based project and a standard GUI screen. The GUI intersperses the user's content with a profusion of visual "widgets" for control: window boundaries, title bars, scroll bars, tool bars, icon trays, view selection buttons, and many more. A whiteboard or project wall, on the other hand, provides a uniform blank surface whose content is the user's marks or posted pieces of paper. To capture the feel of a wall we want to reserve the visual space for content, with an absolute minimum of visual distraction associated with interaction mechanics. We organize our display as an arbitrary collection of opaque and translucent "sheets," which can be created, drawn on, and moved independently. Sheets have only content on them, without visual affordances for actions.

*C. Fluid Interaction*

Along with the visual clutter of GUI interfaces, there are continual interruptions to the flow of activity. Dialog boxes pop up, windows and tools are selected, object handles of various kinds appear and are grabbed, and so on. To some degree this is inevitable to provide complex functionality. The traditional workstation GUI is oriented toward facilitating the speed of highly differentiated actions done by experienced users whose attention focus is entirely on the current computer activity. But users of a wall display are often engaged in simultaneous conversation with people in the room, so their focus on the board is episodic. They do not have the additional cognitive capacity to cope with complex state differences, or have a high tolerance for having their attention distracted to the interaction with the board instead of with the people.

An interactive wall interface needs to provide more functionality than a whiteboard, but this need not call for widgets, modes, dialogs, and the other apparatus of complex interaction.

## VII. DESIGN PROBLEMS

These days high quality HCI Design is difficult to implement because of many reasons [5] .i.e.,
- Market pressure of less time development
- Rapid functionality addition during development
- Excessive several iterations
- Competitive general purpose software
- Expensive errors
- Creativity
- Human behavior analysis

## VIII. CASE STUDY

In this case study we try to limit the software engineering specific issues and we have concentrated on general issues of any interactive system by keeping a general reader in our mind set but some software design specific information is included as a part of our case study especially the sections followed by Class Descriptions, any non software engineer can skip that section as a part of his reading. Case study consists of a process including requirements, finalized requirements, use cases, conceptual class diagrams, class descriptions, physical sketch, identification of GUI design patterns and implemented user interface.

*A. Requirements Specification*

According to the user's requirements, we have to implement a chat application for two more than two user text based communication consisting of some basic but main requirements .i.e.,

- There should be a logging process (user login and logoff.
- Login user can
    - View already online/connected users.
    - Write messages
    - Send that messages to other users
    - Select one user individually or all the users at a time to send or broadcast the messages to all the connected users.
    - Receive and read the responses coming from the other connected users.
    - Attach or upload picture (any) of jpeg format which will be visible to all the connected users.
    - See the picture of selected user in picture window.
    - Picture window should be like optional, I.e. If user doesn't want to see or have picture window then an option is to be provided to close that window.
    - Log off when he wants
    - Quit the application.
- Application should be very simple and interactive.
- Application should be flexible enough so then in future of some changes are needed then can be performed easily without interrupting already implemented stuff.

*B. Finalized Requirements*

From above discussed specified requirements here are some listed finalized functional requirements for HCI Chat application .i.e.,

1. Logging
2. View already connected users
3. Write Message
4. Send message to a particular selected user as well as can broadcast the
5. Message to all connected users
6. Read the responses coming from other users
7. Upload the .jpg picture, which all the connected users can view
8. Also see the other connected user's uploaded picture
9. Can close & open the picture window

*C. Use cases:*

The following illustration shown in Fig 1 is the snapshot of manually designed Use Case diagram of HCI chat application. Below is the list of actions to be performed by the user.

1. Login
2. View Users
3. Write Message
4. Send Message
5. View Response
6. Upload Picture
7. View Uploaded Picture
8. Close / Open Picture Window
9. Logout

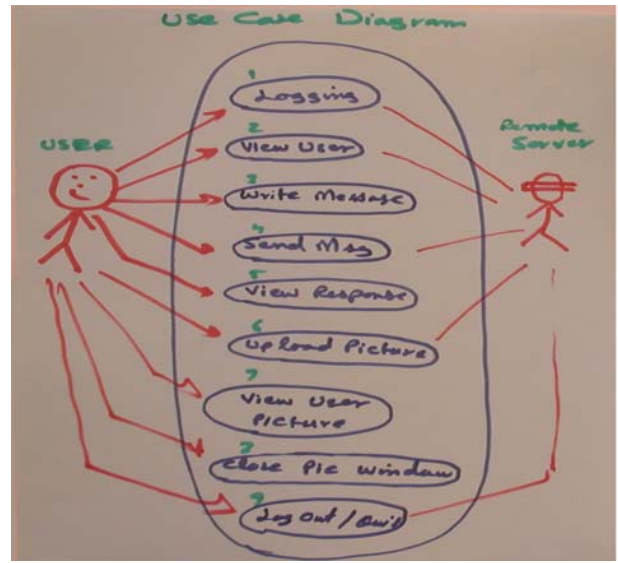

**Fig 1**. Manually deigned Use Case diagram of HCI Chat

*D. Conceptual Diagram*

The following illustration shown in Fig 2 is the snapshot of manually designed Class diagram of HCI chat application's java based classes .i.e., First, Pool, Server, Implementer, PicDialog, UploadImage, Options, AddMenuBar, PicPanel, Listner,

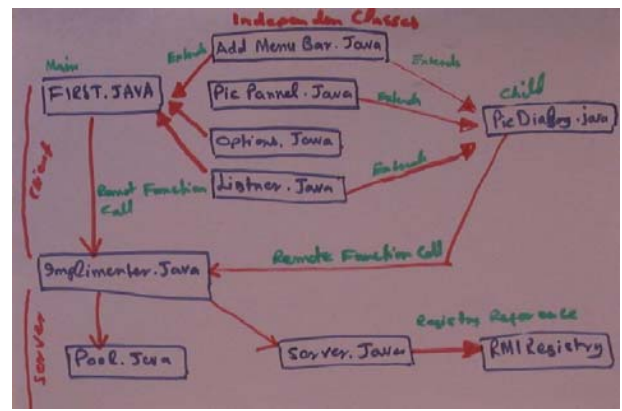

**Fig 2**. Manually deigned Class Diagram of HCI Chat

*1) First. JAVA*

This is the top level class which instantiates other class based on the user requirement. This is the class responsible for generating the parent window responsible for chatting. This class also in turn instantiates many components on the parent window and also spawn the child windows for uploading the pictures of the relevant users of both the ends.

*2) Pool. JAVA*

This is the remote interface which contains the declarations of the function only. This interface is the main interface which provides the information about server's functionality. All the clients are provided with the remote reference to have access to this functionality of the remote Server. In order to access these remote functions every client is provided with the "STUB" and "SKELETON".

*3) Server. JAVA*

This is the remote server class binding the remote object to a "Global Name" in a registry for client access, provides a copy of that registered object to all the clients. Clients can issues a remote call on a function using global name by performing a process called "LOOK UP". The default registry used with this application for binding objects is "RMI REGISTRY".

*4) Implementer. JAVA*

This class provides the definitions for all the functions declared in the "Pool. JAVA". This is the class which actually implements the remote interface and provides the definitions for all the functions declarations in the remote interface. The object bounded in the registry is the object this class. This object is also responsible for exporting the bounded reference. It extends the class "UNICAST REMOTE OBJECT".

*5) PicDialog.JAVA*

This is the window which will display the picture in child window. This window is spawn out by the main window depending upon the user's chosen option. This is also the class which uploads the picture of every user to the server and displays the uploaded image at the connected user's panel. This will also automatically update the panel of the connected user with updated. A dedicated thread is running to monitor automatic change of the picture of the connected users.

*6) UploadImage.JAVA*

Sending & Receiving the images from source and destination. This is the dedicated class running at the client side called by the "Pic Dialog" class when the image of particular user is uploaded or an image of a connected user to be downloaded. In turn the dedicated thread will also continuously accessing this class for downloading the picture from the connected user at regular intervals.

*7) Options. JAVA*

This is an independent class which exclusively provides options to the classes that extends this class. This is designed flexible enough to extend application with new options in future.

*8) AddMenuBar.JAVA*

This is also an independent class which provides the feature 'Menu Bar' with specified menu items to every class that extends this class. Adding Menu bar with specific options of a particular window is very flexible simply by calling a function of this class with new Menu Items. This change will be immediately reflected on the GUI with ease.

*9) PicPanel.JAVA*

This is another independent class which contains panels that could display images on those panels. This class provides any number of panels by simply creating instance of this class. This independent class also provides the panels with picture uploading capability.

*10) Listner.JAVA*

This is the last but most important and independent class which registers the Listener for handling Action Events the application.

### E. Physical Sketch & Implemented GUI

Using the design ideologies, principles, patterns and guild lines we have sketched the graphical user interface of HCI Chat using manual drawing process as shown in Fig 3.

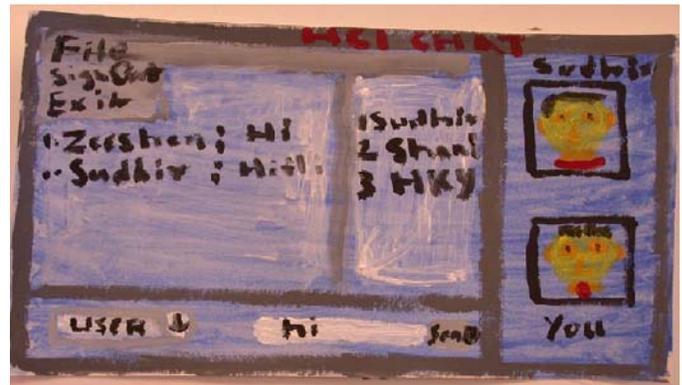

**Fig 3**. Manually deigned HCI Chat Sketch

Then keeping above shown in Fig. 3 manually sketched graphical user interface we designed and implemented real time software graphical user interface of HCI Chat as shown in Fig. 4.

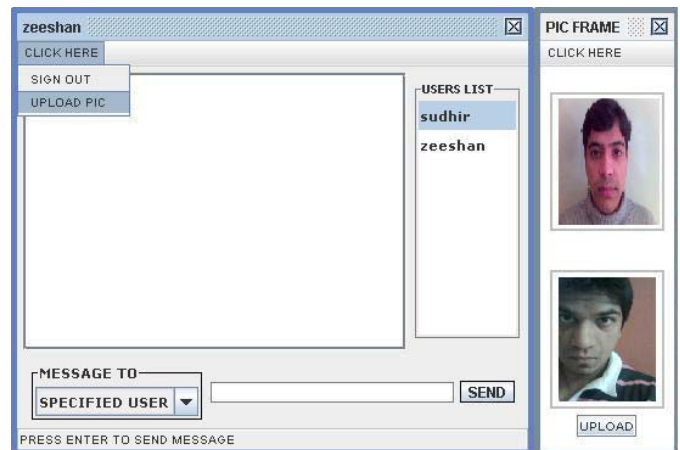

**Fig 4**. Implemented HCI Chat Sketch

### F. Identified GUI Design Patterns in HCI Chat

These are some identified GUI Design Patterns in HCI Chat as shown in Fig .5 which are .i.e.,

*1) Window Per Task*

This pattern is identified in the above figure as a window designed for chat application. This window will also organize all the user interface components required for chatting. There are two different windows which are performing two different tasks (Chat & Picture).

*2) Interaction Style*

This pattern is also identified in the above application as it provides the interaction style for the users in the form of Selection, Direct Manipulation.

*3) Explorable Interface*

This pattern is identified in the above application as it provides an information message to the user before they sign out from the application. This warning is provided because that operation is an undoable operation. Sign out option which asks user, that he really wants to sign out or by mistake he did. More over application will never allow user to select the image other than .jpg and if by mistake he selects other than .jpg file it will prompt him and ask him to select valid one or he has selected the wrong one.

*4) Selection*

This pattern is identified in the above application in the form of single selection. The "Message To" and "List of users" in the above window allows users to select an option among the list of options. Both the selections are single selections.

*5) Direct Manipulation*

This pattern is identified in the above window as the pressing "ENTER Key" allows sending message rather than pressing the "SEND" button on the window, and the "Arrow Keys" allows selecting the options from the "Message To" selection list apart from the mouse selection.

*6) Limited Selection Size*

This pattern is identified in the above window in the case of Group Messaging. To send a group Message definitely the "Message To" options list provides guidance to send group message.

*7) Disabled Irrelevant Things*

This pattern is identified in the above window, as the option "To all Users" from the "Message To" options list is selected then the "Users List" will be disabled. This is to ensure that the message will be sent to all the users in the list but not to a specific user.

*8) Supplementary Windows*

This pattern is identified in Picture Window as well as the dialog window present as the supplementary window.

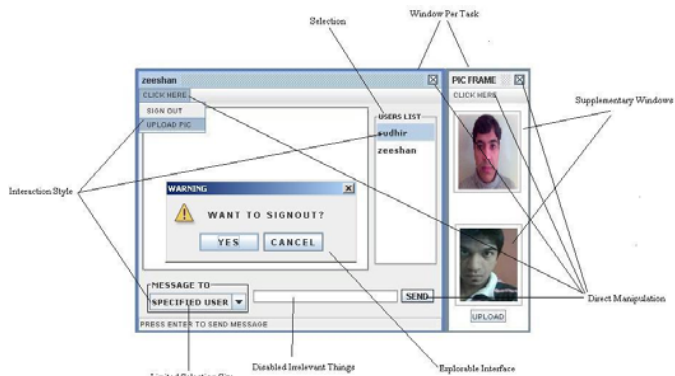

**Fig 5**. Design Pattern Identification in HCI Chat GUI

## IX. CONCLUSION

Our main intention of writing this research paper is to aware graphical user interface designers of almost all the aspects in designing interactive Systems. We have stated our research consisting on the investigation of currently faced major problem to HCI field, brief description of Human Computer Interaction ideologies, design principles, design patterns, design guild lines and basic design goals. Moreover in the end we have presented a case study describing the whole software application development process emphasizing on the design of graphical user interface.